\documentclass{article}
\usepackage{spconf,amsmath,graphicx}
\usepackage{acronym}
\usepackage{url}
\usepackage{subfig,balance}
\usepackage{tabularx}
\usepackage{booktabs,multirow} 
\acrodef{TF}{Time-Frequency}
\acrodef{HTS-AT}{Hierarchical Token-semantic Audio Transformer}
\usepackage{booktabs}
\usepackage{amssymb}

\acrodef{TF}{Time-Frequency}
\acrodef{HTS-AT}{Hierarchical Token-semantic Audio Transformer}

\acrodef{RIR}{Room Impulse Responses}
\acrodef{SER}{Speech Emotion Recognition}
\acrodef{HRI}{Human-Robot Interaction}
\acrodef{CREMA-D}{Crowd-sourced Emotional Multimodal Actors Dataset}

\acrodef{NLP}{Natural Language Processing}
\acrodef{AST}{Audio Spectrogram Transformer}
\acrodef{LSTM}{Long Short-Term Memory}
\acrodef{BLSTM}{Bidirectional Long Short-Term Memory}
\acrodef{RNN}{Recurrent Neural Networks}
\acrodef{CNN}{Convolutional Neural Networks}

\acrodef{ViT}{Vision Transformer}
\acrodef{CLS}{Class Token}
\acrodef{MHA}{multi-head attention}
\acrodef{TCN}{Temporal Convolutional Networks}

\acrodef{ACE}{Acoustic Characterisation of Environments}
\acrodef{CE}{Cross-Entropy}
\acrodef{SOTA}{state-of-the-art}
\acrodef{AST}{Audio Spectrogram Transformer}
\acrodef{ASR}{Automatic Speech Recognition}
\acrodef{BESR}{Binaural Emotional Speech Recognition}
\acrodef{HRI}{Human-Robot Interaction}
\acrodef{SER}{Speech Emotion Recognition}
\acrodef{CNN}{Convolutional Neural Networks}
\acrodef{DNN}{Deep Neural Networks}
\acrodef{RNN}{Recurrent Neural Networks}
\acrodef{BiLSTM}{Bidirectional Long Short-term Memory}
\acrodef{LSTM}{Long Short-term Memory}
\acrodef{MFCC}{mel frequency cepstral coefficients}
\acrodef{RMS}{root mean square}
\acrodef{ZCR}{zero-crossing rate}
\acrodef{CLDNN}{convolutional long- short term deep neural network}
\acrodef{FC}{fully-connected}
\acrodef{FFT}{fast Fourier transform}
\acrodef{STFT}{short-time Fourier transform}
\acrodef{t-SNE}{t-distributed stochastic neighbor embedding}
\acrodef{RAVDESS}{Ryerson audio-visual database of emotional speech and song}
\acrodef{IEMOCAP}{Interactive Emotional Dyadic Motion Capture}

\title{Multi-Microphone Speech Emotion Recognition using the \\ Hierarchical Token-semantic Audio Transformer Architecture}
\name{Ohad Cohen, Gershon Hazan, Sharon Gannot\thanks{The project has received funding from the European Union’s Horizon 2020 Research and Innovation Programme, Grant Agreement No. 871245; and  from the Audition Project, Data Science Program, Council of Higher Education, Israel.}}
\address{Bar-Ilan University, Israel\\
\{ohad.cohen, gershon.hazan, sharon.gannot\}@biu.ac.il
}

\begin{document}
 \ninept
\maketitle
\begin{abstract}
The performance of most emotion recognition systems degrades in real-life situations (“in the wild” scenarios) where the audio is contaminated by reverberation. Our study explores new methods to alleviate the performance degradation of \ac{SER} algorithms and develop a more robust system for adverse conditions. We propose processing multi-microphone signals to address these challenges and improve emotion classification accuracy. We adopt a state-of-the-art transformer model, the \ac{HTS-AT}, to handle multi-channel audio inputs. We evaluate two strategies: averaging mel-spectrograms across channels and summing patch-embedded representations. Our multi-microphone model achieves superior performance compared to single-channel baselines when tested on real-world reverberant environments.


\end{abstract}
\section{Introduction}
\label{sec:intro}
\ac{SER} is widely studied in the literature. Most of the reported studies deal with clean speech data and do not consider additive noise and reverberant environments typical to real-life applications. Only a few studies address the influence of reverberation and noise on \ac{SER}. In these studies, the reverberant data is artificially generated by convolving the clean utterances with \acp{RIR}, either simulated \cite{9980032,Wijayasingha2020RobustnessTN} or recorded in a real environment with various reverberation levels \cite{eyben2013affect,ahmed2017real}. Noise may also be added to the reverberant signals.

A survey of \ac{HRI} in a noisy environment can be found in \cite{zhao2018effect,lollmann2014challenges}. However, this survey does not address the \ac{SER} task. Previous works have shown the significant challenges of detecting emotional speech in large and reverberant rooms. Reverberation can influence the speech signal and negatively affect the predicted results \cite{heracleous2017speech}. The literature on using multiple microphones for \ac{SER} is very scarce. In \cite{bashirpour2018robust}, the robustness of \ac{ASR} systems to emotional speech in noisy conditions is addressed. A \ac{BESR} system is proposed, enabling the simultaneous acquisition of the speaker's emotional state and transcribing the uttered speech signal. 
Devices equipped with multi-microphones are widely available nowadays. Adding the spatial information may improve the performance of audio processing tasks, including \ac{SER}.
However, learning-based algorithms 
are challenged by differences between training and test conditions, specifically a change in the microphone array constellations, e.g., when the number of microphones in train and test conditions is different. 

In early works in the \ac{SER} domain, various architectures such as \ac{CNN}, \ac{DNN}, \ac{RNN}, and \ac{LSTM} were employed. Subsequently, combinations of \ac{CNN} and \ac{RNN} layers emerged, showcasing enhanced performance as compared with traditional classification methodologies \cite{lim2016speech, ma2018emotion, huang2017characterizing}. Additionally, the effectiveness of configurations utilizing blocks comprised of CNN, LSTM and \ac{BiLSTM}, as elucidated in \cite{2022dalia,kwon2020clstm}, was proven effective. Most of the works mainly focus on unimodal learning of emotions, either text, speech, or video \cite{bharti2022text,bhavan2019bagged,abdullah2020facial}. 

The Transformer architecture, initially formulated for \ac{NLP} \cite{9222960, vaswani2017attention}, has found applications in the audio processing domain, including tasks such as speech separation \cite{9413901} and audio classification \cite{gong21b_interspeech}. The superior performance of the \ac{AST} model \cite{gong21b_interspeech}, an adaptation of the \ac{ViT} model \cite{dosovitskiy2020image}, was demonstrated compared to CNN-based models. 
It is imperative to highlight that the \ac{AST} model exclusively handles single-microphone data, whereas, in real-world scenarios, multiple microphones may often be available. A drawback of Transformer models is their reliance on large training data for convergence.

In the current work, we extend the \ac{HTS-AT} \cite{chen2022hts} architecture to accommodate multi-channel inputs, thus enhancing robustness against reverberations. Since only a limited amount of data is available for our task, we resort to fine-tuning already-trained models. Moreover, our scheme can be fine-tuned with a certain number of microphones and tested with another microphone constellation, including microphone positions and the number of microphones. We evaluated the \ac{SER} performance using three datasets, \ac{RAVDESS} \cite{livingstone2018ryerson}, \ac{IEMOCAP} \cite{busso2008iemocap}, and \ac{CREMA-D} \cite{cao2014crema}. We used real-life \acp{RIR} from the \ac{ACE} Challenge  \cite{7486010} to add reverberation to the speech utterances.

\vspace{-0.1cm}
\section{Problem Formulation}
\label{ProblemFormulation }
Let $x(t)$ denote the anechoic signal in the discrete-time domain. An $M$-microphone array captures this signal after propagating in the acoustic enclosure.
The received microphone signals are then given by $y_i(t) = \{x * h_i\}(t), i = 1, 2, \ldots, M$, 
where $h_i(t)$ is the \ac{RIR} from the source position to the position of microphone $i$. This work aims to classify the emotion given the observations $y_i(t);\;i=1,\ldots, M$.

\section{Proposed Model}
\label{section:Proposed_Model}
The proposed \ac{SER} model is based on the Swin-Transformer \cite{liu2021swin},\footnote{\texttt{github.com/microsoft/Swin-Transformer}} a state-of-the-art hierarchical \ac{ViT} \cite{dosovitskiy2020image} architecture, used for a variety of computer vision tasks, that utilizes shifted windows to capture long-range image dependencies. We adopted and modified a Swin Transformer architecture, namely the \acf{HTS-AT} \cite{chen2022hts}, which aims to improve the performance and scalability of audio tasks, such as the AudioSet dataset \cite{gemmeke2017audio}. The \ac{HTS-AT} model is designed to achieve the best performance by reducing the number of parameters, requiring fewer GPU resources, and less training time than the \ac{AST} architecture \cite{chen2022hts}. In the current work, we propose modifying the \ac{HTS-AT} architecture to suit the audio processing requirements better. The main change is to adapt the model to the multi-channel audio processing task, expanding its usefulness beyond the original single-channel design. 
Two alternative multi-channel pre-processing strategies are examined. 
The first strategy applies a summation of patch tokens derived from mel-spectrograms, and the second strategy applies averaging of mel-spectrograms.
Our proposed approach first fuses the multi-channel information into a single stream. Then, the regular \ac{HTS-AT} architecture is applied.

\subsection{Preprocessing and Input Features}
\label{Features}
We assume that the sampling rate of the audio signals is 16~kHz. Each microphone signal is first analyzed by a \ac{STFT}, with a window size of 1024 samples and a hop size of 160 samples. The log-absolute value of the \ac{STFT} bins are then aggregated to construct the mel-spectrograms. 

\subsection{Architecture}
\label{Architecture}
We will now elaborate on the \ac{HTS-AT} architecture. The standard transformer architecture requires extensive computational resources due to the unmodified input token sequence length across all layers. This includes maintaining a large global self-attention matrix and calculating outputs and gradients at each step. The \ac{HTS-AT} architecture is introduced to address these challenges. The \ac{HTS-AT} is designed for supporting multiple audio tasks, e.g., classification, sound event detection, and source localization. It introduces two critical architectural optimizations: a hierarchical structure and a windowed attention mechanism.  
The input audio mel-spectrogram is split into localized patch tokens using a convolutional Patch-Embed \ac{CNN} layer of kernel dimensions $P \times P$, in which the patches are ordered by time segment and frequency bin. Then, tokens propagate through a series of Swin Transformer encoder groups. At the terminus of each group, a Patch Merging layer reshapes the token sequence into its original 2D mel-spectrogram. This layer merges neighboring patches and then embeds them back into a latent space of reduced length. Consequently, the memory requirements decrease exponentially with depth.
Within each Swin Transformer block, window attention is restricted to non-overlapping $W\times W$ squares, partitioning the token sequence. Calculating self-attention within each window subset substantially reduces computational complexity relative to full global attention while capturing localized relationships. As patch size increases downstream, windows encapsulate larger temporal and frequency contexts. Finally, the \ac{HTS-AT} incorporates a token-semantic \ac{CNN} layer after the last transformer block. This layer refines the output by grouping tokens, thus capturing information about their time frames and frequency bins. Consequently, this enhances classification by exploiting token relationships.

\subsection{Multi-Channel Methods}
\label{Single-Channel to Multi-Channel embedding block}
We propose two multi-channel approaches that enable the \ac{HTS-AT} model to handle single- and multi-channel inputs through fine-tuning. Both methods maintain \ac{HTS-AT}'s core architecture to allow flexibility in microphone numbers during fine-tuning and evaluation.
\begin{figure*}[t]
    \centering
      \includegraphics[width=0.92\textwidth]{ser_imeges/Sum.png}
      \caption{Scheme of Patch-Embed Summation.}
      \label{Sum}
   \end{figure*}
   \begin{figure*}[t]
    \centering
      \includegraphics[width=0.92\textwidth]{ser_imeges/Avg_mel.png}
      \caption{Scheme of Average Mel-Spectrograms.}
      \label{Avg}
   \end{figure*}
%

 \noindent\textbf{Patch-Embed Summation:}
Inspired by the study in \cite{Eliav2024CSD}, we embraced a Patch-Embed scheme where each channel undergoes a shared embedding layer followed by summation. This maintains flexibility in the microphone numbers used during fine-tuning and testing without altering the core \ac{HTS-AT} architecture. 
On the left side of Fig.~\ref{Sum}, each of the $M$ channels is analyzed to generate mel-spectrograms, which are then concatenated along the channel depth. SpecAugment \cite{park2019specaugment} is applied collectively across all $M$ mel-spectrograms. Subsequently, each mel-spectrogram is reshaped to the dimensions of a $256 \times 256$ image and passes through the shared Patch-Embed layer. Next, a summation operation is performed across all $M$-encoded channels, making this input suitable for the pre-trained HTS-AT. 
The summation consolidates inter-channel information, enabling more robust representations suitable for reverberant conditions.
The summed embeddings are then fed to the \ac{HTS-AT} feedforward block for the actual emotion classification.

 \noindent\textbf{Average Mel-Spectrograms:}
On the left side of Fig.~\ref{Avg}, the mel-spectrogram of each channel is calculated, followed by an averaging stage that produces a single-channel representation, regardless of the original number of channels. Since reverberation affects each channel differently, the averaged spectrogram will tend to be less reverberant.
%
The averaged mel-spectrogram is then augmented and structured into an image. Subsequently, the image is fed to the Patch-Embed encoder layer, generating a feature tensor with dimensions of $4096 \times 96$.
Similarly to the first architecture, the \ac{HTS-AT} feedforward block implements the emotion classification.
\section{Experimental Study}
\label{Experimental study}
In this section, we describe the experimental setup and present a comparative study of the proposed scheme and a baseline method.
\subsection{Datasets}
Our study comprised three speech emotion recognition datasets, namely \ac{RAVDESS}\cite{livingstone2018ryerson}, \ac{IEMOCAP}\cite{busso2008iemocap} and \ac{CREMA-D}\cite{cao2014crema} datasets.

 The \textbf{\ac{RAVDESS}} dataset comprises 24 actors, evenly distributed between male and female speakers, each uttering 60 English sentences. Hence, there are 1440 utterances expressing eight different emotions: `sad', `happy', `angry', `calm', `fearful', `surprised', `neutral', and `disgust'. All utterances are transcribed in advance. Consequently, emotions are more artificially expressed than in spontaneous
conversation. In this dataset, we decided to merge the emotions `neutral' and `calm' as representations of `neutral'. Therefore, the output of our model is on seven classes instead of eight. A significant drawback of the dataset is its small number of utterances. 

 The \textbf{\ac{IEMOCAP}} dataset comprises approximately 12 hours of speech expressing four emotions: `happy', `sad', `angry' and `neutral'. It consists of conversations between two people that are either improvised or played according to a pre-determined transcript chosen to evoke different emotions.
 
  The \textbf{\ac{CREMA-D}} is a dataset of 7442 original clips from 91 actors comprising 48 male and 43 female actors. Speech utterances were selected from a set of 12 sentences. The sentences were presented using one of six different emotions `anger', `disgust', `fear', `happy', `neutral' and `sad'.

For the multi-channel experiments, we first split the original three datasets as follows: 80\% as a training set, 10\% as a validation set, and 10\% as a test set. We note that the same actor may appear in several splits but not the same utterance. 

We used the 'gpuRIR' Python package\footnote{\texttt{github.com/DavidDiazGuerra/gpuRIR}} to simulate reverberant multi-channel microphone signals (setting the number of microphones to $M=4$) with reverberation time in the range of $T_{60}=200-800$~ms with different room sizes and randomized microphone locations. The room dimensions were uniformly distributed between 3 and 8 meters, with an aspect ratio ranging from 1 to 1.6. The room's height was fixed at 2.9 meters. The sound source was placed at a constant height of 1.75 meters, while the microphones were positioned at a fixed height of 1.6 meters. The $x$ and $y$ coordinates were randomly determined within the room for both the sound source and microphones, while keeping them at least 0.5 meters away from the room walls. By doing so, we were also able to enlarge the datasets. For \ac{RAVDESS}, 6863 reverberant speech samples were generated for training, and 852 samples were used for validation. We follow the same procedure for \ac{IEMOCAP} with 7356 train samples and 2107 validation samples and for \ac{CREMA-D} with 5945 train samples and 1487 validation samples. Our objective encompasses evaluating our model in real-world reverberant environments. To achieve this, we conducted tests utilizing the \ac{ACE} \ac{RIR} database \cite{7486010}, which comprises seven distinct rooms characterized by varying dimensions and exhibiting diverse ranges of $T_{60}$. We used the subset captured by a mobile phone equipped with three microphones, which are frequently used for speech interaction, making them a practical choice for real-world \ac{SER} applications. The model was fine-tuned with the synthesized \acp{RIR} and evaluated with the test sets of the various datasets convolved with the \ac{ACE} \acp{RIR}. 
\begin{table}[t]
    \caption{The results for the non-reverberant single-microphone case. The fine-tuned \ac{HTS-AT} model is compared with \cite{2022dalia}. We present the weighted average accuracy (in percent) of 20 models that achieved the highest accuracy on the validation set and tested on the test set.}
    \label{tab:Single-Channel fine-tuning results}
    \centering
    \resizebox{0.99\columnwidth}{!}{
\begin{tabular}{ lccc }
 \toprule
  Models & RAVDESS & IEMOCAP& CREMA-D \\
  \midrule
 Fine-tuned HTS-AT  & \textbf{90}\%   &\textbf{70.93}\% &\textbf{75.86}\%\\
 \ac{BiLSTM} + Attention\cite{2022dalia} & 80\%   &66\%  & - \\
 \bottomrule
\end{tabular}
}
\vspace{-0.1cm}
\end{table}
\begin{table*}[t]
  \renewcommand*{\arraystretch}{1.1}
  \centering
    \caption{Results on the test sets of the RAVDESS, IEMOCAP, and CREMA-D datasets convolved with \acp{RIR} of three microphones from the \ac{ACE}  database. The `HTS-AT' columns are fine-tuned on reverberant single-channel audio. The `Avg mel' columns depict results where mel-spectrograms were averaged across four channels during fine-tuning and tested on three channels. The `Sum PE' columns are the Patch-Embed Summation approach fine-tuned and tested on three channels.}
%
    \resizebox{2.05\columnwidth}{!}{
    \begin{tabular}{l ccc ccc ccc}
      \toprule
      \multirow{ 2}{*}{Room ($T_{60}$~[ms])} &  \multicolumn{3}{c}{RAVDESS} & \multicolumn{3}{c}{IEMOCAP} & \multicolumn{3}{c}{CREMA-D} \\
      \cmidrule(lr){2-4}\cmidrule(r){5-7} \cmidrule(r){8-10}
      & HTS-AT & Avg mel & Sum PE & HTS-AT & Avg mel & Sum PE & HTS-AT & Avg mel & Sum PE  \\
      \midrule
      Lecture  1 ($638$) & 77.3 (70.6-84.0) & 80.6 (74.6-86.6) & 81.3 (74.6-87.3) & 61.3 (57.1-65.2) & 67.0 (63.0-71.0) & 67.4 (63.6-71.2) & 63.2 (60.1-66.6) & 66.4 (63.2-70.1) & 67.4 (64.2-71.0)  \\
      Lecture  2  ($1220$) & 77.3 (70.6-83.3) & 78.6 (71.3-84.6) & 82.0 (76.0-88.0) & 63.4 (59.5-67.4) & 66.1 (62.4-70.0) & 68.7 (65.0-72.4) & 65.4 (62.0-68.9) & 67.3 (64.0-70.8) & 66.0 (62.5-69.3)   \\
      Lobby  ($646$) & 78.6 (72.0-84.6) & 82.0 (76.0-88.0) & 84.0 (78.0-89.3) & 61.8 (58.0-65.8) & 64.0 (60.0-67.8) & 65.1 (61.3-68.8) & 64.2 (60.9-67.8) & 66.2 (62.8-69.8) & 66.8 (63.3-70.2)  \\
      Meeting  1 ($437$) & 73.3 (66.6-80.0) & 80.6 (74.0-86.6) & 82.6 (76.6-88.0) & 62.9 (58.9-66.9) & 66.7 (62.9-70.6) & 68.7 (65.1-72.4)  & 64.4 (61.0-67.8) & 65.7 (62.5-69.3) & 65.7 (62.2-69.0)    \\
      Meeting  2 ($371$) & 82.0 (76.0-87.3) & 83.3 (76.6-89.3) & 85.3 (80.0-90.6) & 59.5 (55.7-63.6) & 66.3 (62.4-70.3) & 64.2 (60.2-68.1) & 65.6 (62.4-69.0) & 66.6 (63.4-69.8) & 66.8 (63.6-70.2)  \\
      Office 1 ($332$) & 76.0 (69.3-82.6) & 83.3 (77.3-88.6) & 80.6 (74.0-86.6) & 63.6 (59.7-67.6) & 68.3 (64.7-72.3)  & 67.2 (63.6-71.0) & 64.1 (60.1-67.6) & 68.8 (65.4-72.4) & 67.3 (64.0-71.0)         \\
      Office 2 ($390$) & 78.6 (72.6-84.6) & 81.3 (75.3-87.3) & 82.6 (76.0-88.0) & 59.5 (55.5-63.8) & 64.7 (61.0-68.8)  & 65.4 (61.5-69.6) & 62.6 (59.4-66.1) & 64.0 (60.6-67.6) & 64.1 (60.6-67.7)       \\
        
      \bottomrule
       \label{ACE_results}
    \end{tabular}
    }
\end{table*}

\subsection{Algorithm Setup}
\label{Setup}
We fine-tuned our model starting from the original \ac{HTS-AT} model that was pre-trained utilizing the AudioSet dataset, which comprises over two million audio samples. Each sample is 10 seconds long and is categorized into 527 distinct sound event categories. This pre-training provides a strong foundation for audio feature extraction, which we then adapt to our specific \ac{SER}. In order to adapt the pre-trained AduioSet model to \ac{RAVDESS}, \ac{IEMOCAP}, and \ac{CREMA-D}, we adjusted the number of output classes to 7, 4, and 6, respectively. 
In the three datasets, pre-processing and warm-up strategy were carried out as demonstrated in \cite{chen2022hts} by providing the \ac{HTS-AT} with 64 mel-bins to compute the \ac{STFT} and mel-spectrograms features with 160 hop size and 1024 window size. We modified the original AdamW optimizer of \ac{HTS-AT} to the traditional Adam optimizer with a learning rate of $1e^{-3}$, and batch size of 128. We set the number of epochs to 150 for all datasets. We used cross-entropy loss as the metric. We also follow the same original \ac{HTS-AT} hyperparameter settings. The patch dimensions are set to $4 \times 4$, the patch window length is 256 frames, and the attention window size is $8 \times 8$. The architectural configuration comprises four network groups, each comprising 2, 2, 6, and 2 Swin-Transformer blocks, respectively. The initial patch embed is linearly projected to a dimension of $D=96$, and correspondingly, after each transformer group, the dimension increases exponentially to $8D=768$, aligning seamlessly with the principles of \ac{AST}. In part of our experiments, we reduced the depth of each network group by half, thereby reducing the number of trainable parameters by half to prevent overfitting. In addition, we added an early stopping strategy with a patience of 50 for RAVDESS and 25 for IEMOCAP and CREMA-D. The overall number of parameters for the fine-tuned \ac{HTS-AT} was 15.7M for the RAVDESS and 28.6M for IEMOCAP and CREMA-D. Since only fine-tuning is required, we could train the models in less than two hours on an A6000 RTX GPU.

\subsection{Results}
\label{Results}
We begin our experimental study with the single-microphone case. We compare the fine-tuned \ac{HTS-AT} model with another model developed in our group that utilizes \ac{BiLSTM} combined with an Attention mechanism \cite{2022dalia}, which has achieved very good results on the \ac{RAVDESS} and \ac{IEMOCAP} datasets. We applied both schemes to a non-reverberant single-microphone signal.
The results are presented in Table~\ref{tab:Single-Channel fine-tuning results}. 

Furthermore, to complete establishing the baseline using the \ac{RAVDESS} and \ac{IEMOCAP} datasets, we fine-tuned another single-channel \ac{HTS-AT} model by fine-tuning on the artificially reverberated datasets with a uniformly distributed $T_{60}=200-800$~ms. Then, we compared the three models by computing the Accuracy with statistical significance measure.\footnote{\texttt{github.com/luferrer/ConfidenceIntervals}} The evaluation was performed on the reverberant test set using the \ac{ACE} database only applied to one of the microphones. 
\begin{figure}[ht]
    \centering
    \subfloat[\centering IEMOCAP]{{\includegraphics[width=0.52\columnwidth]{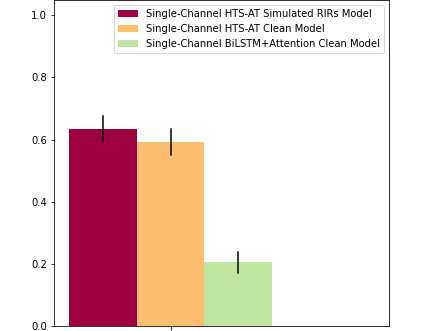} }}
    \subfloat[\centering RAVDESS]{{\includegraphics[width=0.51\columnwidth]{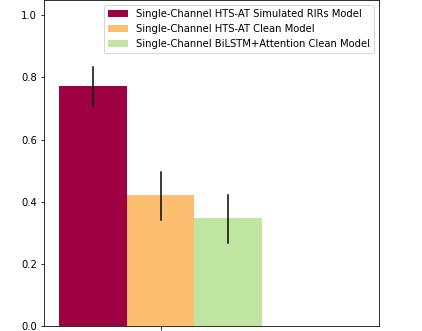} }}
        \centering
    \caption{Accuracy and Confidence Interval on test sets convolved with ACE RIR Lecture Room 2 ($T_{60}=1220~ms$). The results of two \ac{HTS-AT} fine-tuned on either clean or simulated RIRs datasets compared with \cite{2022dalia} trained on the clean datasets.}
    \label{HTS-AT_VS_BLSTM}
\end{figure}
Examining Fig.~\ref{HTS-AT_VS_BLSTM} for both the \ac{IEMOCAP} and \ac{RAVDESS} datasets, it is evident that the two \ac{HTS-AT} variants outperform the approach in \cite{2022dalia}. It is also clear that for \ac{RAVDESS}, the \ac{HTS-AT} fine-tuned with reverberant speech significantly outperforms the \ac{HTS-AT} fine-tuned using clean data. While this is also true for \ac{IEMOCAP}, the differences are less significant. In addition, the significant gap observed for the RAVDESS dataset compared to IEMOCAP is due to differences in the nature of the datasets. RAVDESS contains more controlled, acted emotions, while IEMOCAP is more naturalistic.

We now turn to the evaluation of the multi-channel schemes. 
In all experiments, we tested all three datasets with the utterances convolved with the three \acp{RIR} of the \ac{ACE} database. We only report the results for the remote source case to emphasize the reverberation effects. 
The average mel-spectrogram scheme was fine-tuned using four microphones. The Patch-Embed Summation scheme was fine-tuned using three microphones. While we used different microphone counts ($M=3$ and $M=4$) for fine-tuning, this was done to demonstrate the flexibility of our approach in handling varying numbers of input channels. However, we acknowledge that using the same number of microphones for all models would provide a more direct comparison. In both cases, the \acp{RIR} were generated using the simulator, as explained above. All experimental procedures followed identical training-validation splits, maintaining consistent model configurations, sizes, and hyperparameters for each dataset.

Table~\ref{ACE_results} provides a comparative assessment of three model variants, all fine-tuned with reverberant speech: 1) `HTS-AT' - a single-channel scheme, 2) `Avg mel' - mel-spectrograms averaged across four channels, and 3) `Sum PE' - Patch-Embed Summation across three channels. 
In the table, we report the accuracy and the associated confidence interval for all seven rooms and three databases.
Examining these results indicates the consistent, albeit modest, improvements achieved by the `Sum PE' approach. Embedding each channel separately before summing, the model captures subtle inter-channel differences that might be lost in averaging. This allows the model to emphasize the most informative aspects of each channel for emotion recognition, leading to improved performance. Importantly, these benefits come without a considerable increase in computational complexity compared to single-channel models. 

\section{Conclusions}
\label{Conclusions}
In this paper, we presented a multi-microphone transformer-based model for \ac{SER} in reverberant environments. Based on the HTS-AT architecture, the model employs two strategies for handling multi-channel audio inputs: averaging mel-spectrograms across channels and summing patch-embedded representations. When tested on real-world reverberant environments, the results show improved \ac{SER} accuracy compared to single-channel schemes. By leveraging spatial information from multiple microphones, our model was able to exhibit a more robust behavior of the \ac{SER} in challenging acoustic conditions. The consistent, but not large, improvements of the proposed multi-microphone schemes show promise for developing \ac{SER} systems that can perform reliably in noisy and reverberant scenarios.
\balance
\bibliographystyle{IEEEtran}

\bibliography{template}

\begin{thebibliography}{10}
\providecommand{\url}[1]{#1}
\csname url@samestyle\endcsname
\providecommand{\newblock}{\relax}
\providecommand{\bibinfo}[2]{#2}
\providecommand{\BIBentrySTDinterwordspacing}{\spaceskip=0pt\relax}
\providecommand{\BIBentryALTinterwordstretchfactor}{4}
\providecommand{\BIBentryALTinterwordspacing}{\spaceskip=\fontdimen2\font plus
\BIBentryALTinterwordstretchfactor\fontdimen3\font minus
  \fontdimen4\font\relax}
\providecommand{\BIBforeignlanguage}[2]{{%
\expandafter\ifx\csname l@#1\endcsname\relax
\typeout{** WARNING: IEEEtran.bst: No hyphenation pattern has been}%
\typeout{** loaded for the language `#1'. Using the pattern for}%
\typeout{** the default language instead.}%
\else
\language=\csname l@#1\endcsname
\fi
#2}}
\providecommand{\BIBdecl}{\relax}
\BIBdecl

\bibitem{9980032}
T.~Guo, S.~Li, M.~Unoki, and S.~Okada, ``Investigation of
  noise-reverberation-robustness of modulation spectral features for
  speech-emotion recognition,'' in \emph{Asia-Pacific Signal and Information
  Processing (APSIPA)}, 2022, pp. 39--46.

\bibitem{Wijayasingha2020RobustnessTN}
L.~N.~S. Wijayasingha and J.~A. Stankovic, ``Robustness to noise for speech
  emotion classification using {CNNs} and attention mechanisms,'' \emph{Smart
  Health}, 2020.

\bibitem{eyben2013affect}
F.~Eyben, F.~Weninger, and B.~Schuller, ``Affect recognition in real-life
  acoustic conditions: a new perspective on feature selection,'' in
  \emph{INTERSPEECH}, Lyon, France, 2013.

\bibitem{ahmed2017real}
M.~Y. Ahmed, Z.~Chen, E.~Fass, and J.~Stankovic, ``Real time distant speech
  emotion recognition in indoor environments,'' in \emph{Int. Conf. on Mobile
  and Ubiquitous Systems: Computing, Networking and Services}, 2017, pp.
  215--224.

\bibitem{zhao2018effect}
S.~Zhao, Y.~Yang, and J.~Chen, ``Effect of reverberation in speech-based
  emotion recognition,'' in \emph{IEEE Int. Conf. on the Science of Electrical
  Engineering in Israel (ICSEE)}, 2018.

\bibitem{lollmann2014challenges}
H.~W. L{\"o}llmann, H.~Barfuss, A.~Deleforge, S.~Meier, and W.~Kellermann,
  ``Challenges in acoustic signal enhancement for human-robot communication,''
  in \emph{Speech Communication; ITG Symposium}, 2014.

\bibitem{heracleous2017speech}
P.~Heracleous, K.~Yasuda, F.~Sugaya, A.~Yoneyama, and M.~Hashimoto, ``Speech
  emotion recognition in noisy and reverberant environments,'' in \emph{Int.
  Conf. on Affective Computing and Intelligent Interaction (ACII)}, 2017, pp.
  262--266.

\bibitem{bashirpour2018robust}
M.~Bashirpour and M.~Geravanchizadeh, ``Robust emotional speech recognition
  based on binaural model and emotional auditory mask in noisy environments,''
  \emph{EURASIP J. on Audio, Speech, and Music Processing}, vol. 2018, no.~1,
  pp. 1--13, 2018.

\bibitem{lim2016speech}
W.~Lim, D.~Jang, and T.~Lee, ``Speech emotion recognition using convolutional
  and recurrent neural networks,'' in \emph{Asia-Pacific Signal and Information
  Processing (APSIPA)}, 2016.

\bibitem{ma2018emotion}
X.~Ma, Z.~Wu, J.~Jia, M.~Xu, H.~Meng, and L.~Cai, ``Emotion recognition from
  variable-length speech segments using deep learning on spectrograms.'' in
  \emph{Interspeech}, 2018, pp. 3683--3687.

\bibitem{huang2017characterizing}
C.-W. Huang, S.~Narayanan \emph{et~al.}, ``Characterizing types of convolution
  in deep convolutional recurrent neural networks for robust speech emotion
  recognition,'' \emph{arXiv preprint arXiv:1706.02901}, 2017.

\bibitem{2022dalia}
D.~Sherman, G.~Hazan, and S.~Gannot, ``Study of speech emotion recognition
  using {BLSTM} with attention,'' in \emph{European Signal Processing Conf.
  (EUSIPCO)}, Helsinki, Finland, Sep. 2023.

\bibitem{kwon2020clstm}
S.~Kwon \emph{et~al.}, ``{CLSTM: D}eep feature-based speech emotion recognition
  using the hierarchical {ConvLSTM} network,'' \emph{Mathematics}, vol.~8,
  no.~12, p. 2133, 2020.

\bibitem{bharti2022text}
S.~K. Bharti, S.~Varadhaganapathy, R.~K. Gupta, P.~K. Shukla, M.~Bouye, S.~K.
  Hingaa, and A.~Mahmoud, ``Text-based emotion recognition using deep learning
  approach,'' \emph{Computational Intelligence and Neuroscience}, vol. 2022,
  no.~1, p. 2645381, 2022.

\bibitem{bhavan2019bagged}
A.~Bhavan, P.~Chauhan, R.~R. Shah \emph{et~al.}, ``Bagged support vector
  machines for emotion recognition from speech,'' \emph{Knowledge-Based
  Systems}, vol. 184, p. 104886, 2019.

\bibitem{abdullah2020facial}
M.~Abdullah, M.~Ahmad, and D.~Han, ``Facial expression recognition in videos:
  An cnn-lstm based model for video classification,'' in \emph{Int. Conf. on
  Electronics, Information, and Communication (ICEIC)}, 2020.

\bibitem{9222960}
A.~Gillioz, J.~Casas, E.~Mugellini, and O.~A. Khaled, ``Overview of the
  transformer-based models for {NLP} tasks,'' in \emph{Conf. on Computer
  Science and Information Systems (FedCSIS)}, 2020, pp. 179--183.

\bibitem{vaswani2017attention}
A.~Vaswani, N.~Shazeer, N.~Parmar, J.~Uszkoreit, L.~Jones, A.~N. Gomez,
  {\L}.~Kaiser, and I.~Polosukhin, ``Attention is all you need,''
  \emph{Advances in neural information processing systems}, vol.~30, 2017.

\bibitem{9413901}
C.~Subakan, M.~Ravanelli, S.~Cornell, M.~Bronzi, and J.~Zhong, ``Attention is
  all you need in speech separation,'' in \emph{IEEE Int. Conf. on Acoustics,
  Speech and Signal Processing (ICASSP)}, 2021, pp. 21--25.

\bibitem{gong21b_interspeech}
Y.~Gong, Y.-A. Chung, and J.~Glass, ``{AST: A}udio spectrogram transformer,''
  in \emph{Interspeech}, 2021, pp. 571--575.

\bibitem{dosovitskiy2020image}
A.~Dosovitskiy, L.~Beyer, A.~Kolesnikov, D.~Weissenborn, X.~Zhai,
  T.~Unterthiner, M.~Dehghani, M.~Minderer, G.~Heigold, S.~Gelly \emph{et~al.},
  ``An image is worth 16x16 words: {T}ransformers for image recognition at
  scale,'' in \emph{Int. Conf. on Learning Representations (ICLR)}, 2021.

\bibitem{chen2022hts}
K.~Chen, X.~Du, B.~Zhu, Z.~Ma, T.~Berg-Kirkpatrick, and S.~Dubnov, ``Hts-at: A
  hierarchical token-semantic audio transformer for sound classification and
  detection,'' in \emph{IEEE Int. Conf. on Acoustics, Speech and Signal
  Processing (ICASSP)}, 2022, pp. 646--650.

\bibitem{livingstone2018ryerson}
S.~R. Livingstone and F.~A. Russo, ``The {Ryerson} audio-visual database of
  emotional speech and song ({RAVDESS}): A dynamic, multimodal set of facial
  and vocal expressions in north american english,'' \emph{PloS one}, vol.~13,
  no.~5, p. e0196391, 2018.

\bibitem{busso2008iemocap}
C.~Busso, M.~Bulut, C.-C. Lee, A.~Kazemzadeh, E.~Mower, S.~Kim, J.~N. Chang,
  S.~Lee, and S.~S. Narayanan, ``{IEMOCAP}: Interactive emotional dyadic motion
  capture database,'' \emph{Language resources and evaluation}, vol.~42, no.~4,
  pp. 335--359, 2008.

\bibitem{cao2014crema}
H.~Cao, D.~G. Cooper, M.~K. Keutmann, R.~C. Gur, A.~Nenkova, and R.~Verma,
  ``Crema-d: Crowd-sourced emotional multimodal actors dataset,'' \emph{IEEE
  Trans. on affective computing}, vol.~5, no.~4, pp. 377--390, 2014.

\bibitem{7486010}
J.~Eaton, N.~D. Gaubitch, A.~H. Moore, and P.~A. Naylor, ``Estimation of room
  acoustic parameters: {T}he {ACE} challenge,'' \emph{IEEE/ACM Trans. on Audio,
  Speech, and Language Processing}, vol.~24, no.~10, pp. 1681--1693, 2016.

\bibitem{liu2021swin}
Z.~Liu, Y.~Lin, Y.~Cao, H.~Hu, Y.~Wei, Z.~Zhang, S.~Lin, and B.~Guo, ``Swin
  transformer: Hierarchical vision transformer using shifted windows,'' in
  \emph{IEEE/CVF Int. Conf. on computer vision}, 2021, pp. 10\,012--10\,022.

\bibitem{gemmeke2017audio}
J.~F. Gemmeke, D.~P. Ellis, D.~Freedman, A.~Jansen, W.~Lawrence, R.~C. Moore,
  M.~Plakal, and M.~Ritter, ``Audio set: {A}n ontology and human-labeled
  dataset for audio events,'' in \emph{IEEE Int. Conf. on acoustics, speech and
  signal processing (ICASSP)}, 2017, pp. 776--780.

\bibitem{Eliav2024CSD}
A.~Eliav and S.~Gannot, ``Concurrent speaker detection: A multi-microphone
  transformer-based approach,'' in \emph{European Signal Processing Conf.
  (EUSIPCO)}, France, Aug. 2024.

\bibitem{park2019specaugment}
D.~S. Park, W.~Chan, Y.~Zhang, C.-C. Chiu, B.~Zoph, E.~D. Cubuk, and Q.~V. Le,
  ``Specaugment: A simple data augmentation method for automatic speech
  recognition,'' \emph{arXiv preprint arXiv:1904.08779}, 2019.

\end{thebibliography}

\end{document}